\documentclass[10pt,twocolumn,floatfix,aps,prl,showpacs,groupedaddress]{revtex4}
\usepackage{epsfig}
\usepackage{float,amssymb,amsmath}
\usepackage{color}
\DeclareGraphicsExtensions{.pdf,.png,.jpg}
\begin{document}

\title{Anisotropic transport and magnetic properties, and magnetic-field tuned states of CeZn$_{11}$ single crystals.}

\author{H. Hodovanets$^{1,2}$, S. L. Bud'ko$^{1,2}$, X. Lin$^2$, V. Taufour$^2$, M. G. Kim$^{1,2}$\footnote{Currently at the Lawrence Berkeley National Laboratory (LBL)}, D. K. Pratt$^{1,2}$\footnote{Currently at the  National Institute of Standards and Technology (NIST)}, A. Kreyssig$^{1,2}$, and P. C. Canfield$^{1,2}$}

\affiliation{$^1$ The Ames Laboratory U.S. DOE Iowa State University Ames, IA 50011, USA}
\affiliation{$^2$ Department of Physics \& Astronomy, Iowa State University, Ames, IA 50011, USA}

\begin{abstract}
We present detailed temperature and field dependent data obtained from magnetization, resistivity, heat capacity, Hall resistivity and thermoelectric power measurements performed on single crystals of CeZn$_{11}$. The compounds orders antiferromagnetically at $\sim$ 2 K. 
The zero-field resistivity and TEP data show features characteristic of a Ce-based intermetallic with crystal electric field splitting and possible correlated, Kondo lattice effects.
We constructed the $T-H$ phase diagram for the magnetic field applied along the easy, [110], direction which shows that the magnetic field required to suppress $T_N$ below 0.4 K is in the range of 45-47.5 kOe. 
A linear behavior of the $\rho(T)$ data, $\bf{H}\|$[110], was observed only for $H$=45 kOe for 0.46 K$\leq T\leq$1.96 K followed by the Landau-Fermi-liquid regime for a limited range of fields, 47.5 kOe$\leq H\leq$60 kOe. From the analysis of our data, it appears that CeZn$_{11}$ is a weakly to moderately correlated local moment compound with rather small Kondo temperature. The thermoelectric and transport properties of CeZn$_{11}$ are mostly governed by the CEF effects. Given the very high quality of our single crystals, quantum oscillations are found for both CeZn$_{11}$ and its non-magnetic analogue, LaZn$_{11}$.
\end{abstract}


\maketitle

\section{Introduction}
The broad interest in intermetallic compounds containing Ce is due, in part, to the fact that these compounds sometimes show anomalous electronic and magnetic properties associated with heavy fermions or valence-fluctuation effects \cite{Stewart1984,Stewart2001, Stewart2006, Stockert2011}. A competition between local moment ordering mediated via the Ruderman-Kittel-Kasuya-Yosida (RKKY) interaction and the on-site Kondo fluctuation of the $f$ electrons leads to the variety of the ground states in these compounds. In addition, the crystal electric field (CEF) splitting of the Hunds rule ground state multiplet, $J$, influences the temperature-dependent thermodynamic and transport properties.

With this in mind, polycrytalline samples of CeZn$_{11}$ were investigated by susceptibility and specific heat measurements for possible heavy-electron behavior \cite{Nakazawa1993}. It turned out that CeZn$_{11}$ was reported to have a relatively small $\gamma$ value of 40 mJ/(mol K$^2$) which was hard to estimate precisely because of the antiferromagnetic (AFM) transition near 2.0 K. It is precisely this relatively low value of the N\'{e}el temperature, $T_N$, that makes CeZn$_{11}$ a promising candidate for the study of magnetic field tuning and possible quantum critical point (QCP) effects. 

CeZn$_{11}$ crystallizes in tetragonal, $\it I$4$_1/amd$, BaCd$_{11}$-type structure. The unit cell contains one unique Ce site with tetragonal point symmetry, and three crystallographically distinct Zn sites. The Ce atom occupies the center of a polyhedron composed with 22 zinc atoms \cite{Zelinska2004}. 

In order to gain insight into the anisotropic, low temperature physical properties of  CeZn$_{11}$ and investigate the possibility of a magnetic field induced QCP, we grew single crystals and performed transport and thermodynamic measurements down to 0.4 K with applied magnetic fields of up to 140 kOe. We present here anisotropic temperature and field dependent magnetic susceptibility, resistivity, Hall effect, heat capacity and thermoelectric power (TEP) measurements on CeZn$_{11}$. From these measurements we assembled a $\it T-H$ phase diagram that shows the evolution of the magnetic field induced states of CeZn$_{11}$. 

\section{Experimental}

Single crystals of CeZn$_{11}$, in the form of slightly distorted octahedra, were grown from high temperature binary solutions rich in Zn. \cite{Canfield2010,Canfield1992} High purity, elemental Ce (The Ames Laboratory) and Zn (5N, Alfa Aesar) were combined in an alumina crucible in the molar ratio of 1.5:98.5 respectively and sealed in a silica ampule under a partial pressure of high purity argon gas. The ampule, in a 50 ml alumina crucible, placed on $\sim$ 3 cm alumina slab, was heated to 1000 $^0$C, held there for 3 h, then cooled over 3 h to 850 $^0$C, and finally cooled down over 100 h to 500 $^0$C, at which temperature the excess Zn was decanted using a $g$-enhancing rotational method. \cite{Canfield2010,Canfield1992} Although the single crystals obtained had some of the reflective surfaces covered in Zn flux, the residual flux and/or oxide slag on the crystal surface could be removed by using the dilute acid (0.5 vol $\%$ of HCl in H$_2$O) \cite{Jia2009} or polished off as in the case of samples for resistivity, Hall effect and thermoelectric power measurements.

LaZn$_{11}$ single crystals are useful since they can provide an estimate of the non-magnetic contribution to the transport and thermodynamic measurements of CeZn$_{11}$. Although CeZn$_{11}$ is the most Zn rich Ce-Zn binary compound \cite{Massalski1990}, and as such can be grown out of excess Zn readily, Ref. \cite{Berche2009} indicates that LaZn$_{11}$ is bracketed on either side by the more Zn-rich LaZn$_{13}$ and the less Zn-rich La$_2$Zn$_{17}$. This binary phase diagram, if accurate, would imply single crystal growth of LaZn$_{11}$ would be difficult. On the other hand, according to Refs. \cite{Berche2012,Okamoto2011}, a substantial liquidus line for LaZn$_{11}$ exists and single crystals should be easily grown. To grow single crystals of LaZn$_{11}$, the high purity elemental La (The Ames Laboratory) and Zn (5N, Alfa Aesar) were put in the molar ratio of 3:97 respectively into an alumina crucible then sealed in a silica ampule under the partial pressure of a high purity argon gas. The ampule, in a 50 ml alumina crucible, placed on $\sim$3 cm alumina slab, was heated to 950 $^0$C, held there for 2 h, in 1 h cooled to 870 $^0$C held there for 1 h and then cooled over 75 h to 720 $^0$C, at which temperature the excess Zn was decanted. Single crystals of LaZn$_{11}$ obtained were very similar in the morphology to those of CeZn$_{11}$. This result confirms the extended liquidus line shown for LaZn$_{11}$ in Refs. \cite{Berche2012,Okamoto2011}. 

Powder x-ray diffraction (XRD) data were collected on a Rigaku MiniFlex diffractometer (Cu $K_{\alpha 1,2}$ radiation) at room temperature. Lattice parameters were refined by the LeBail method using Rietica software \cite{rietica}. Laue-back-reflection patterns were taken with a MWL-110 camera manufactured by Multiwire Laboratories.

Magnetic measurements were carried out in a Quantum Design, Magnetic Property Measurement System (MPMS), SQUID magnetometer. For the magnetization measurements, the sample was glued with Loctite 495 glue to a Kel-F$^{\textcircled{R}}$ PCTFE (PolyChloroTriFluoroEthylene) disk, that had been machined so as to tightly fit inside of a transparent plastic straw used as a sample holder for magnetization measurements \cite{MPMS}, with the c-axis being perpendicular to the disk. To apply the field parallel to the [100] or the [110] direction the disk with the sample were mounted vertically in between two straws. Although this measuring protocol allowed for ready orientation of the sample, it also made it hard to have alignments more accurate than $\pm$10$^0$ of desired direction of the applied field. The magnetic signal from disk with the glue was accounted for in the final results.

The temperature- and field-dependent resistivity, Hall resistivity and heat capacity measurements were performed in applied fields up to 140 kOe and temperatures down to $\sim $0.4 K in a Quantum Design Physical Property Measurement System (PPMS-14) with a He-3 option. A standard 4-probe geometry, $\it ac$ technique ($\it f$=16 Hz, $\it I$=3-0.3 mA), was used to measure the electrical resistance of the samples. Electrical contact to the samples was made with platinum wires attached to the samples using EpoTek H20E silver epoxy. To calculate the resistivity of two samples of CeZn$_{11}$ and a sample of LaZn$_{11}$, the cross-sectional area and the distance between the midpoints of the two voltage contacts was used. For the two directions of current flow for CeZn$_{11}$, $\it l$=0.69$\pm$0.01 mm for $\bf{I}\|$[001] and $\it l$=0.94$\pm$0.01 mm for $\bf{I}\|$[010], the width of both voltage contact together was 0.27$\pm$0.01 mm for $\bf{I}\|$[001] and 0.26$\pm$0.01 mm for $\bf{I}\|$[010]. The virtually equal resistivity values for both current orientations is most likely a coincidence due to the chosen criterion of estimation of the distance between two voltage contacts. If the distance between the outside and inside ends of contacts is used then the value of resistivity will be 28$\%$ smaller and 64 $\%$ larger ($\bf{I}\|$[001]) respectively, and 22$\%$ smaller and 38$\%$ larger ($\bf{I}\|$[010]) respectively of that calculated for the criterion chosen above. We believe that in the present case, the larger error in the calculating of the resistivity comes from estimating the distance between the voltage contacts rather than from measurements of the cross-section area of the samples. The cross-section areas (W$\times$H) of the samples used in this work were $0.28\times$ 0.12 mm$^2$ ($\bf{I}\|$[001]) and $0.17\times$ 0.08 mm$^2$ ($\bf{I}\|$[010]).  

The same criterion was used to determine the resistivity of LaZn$_{11}$ single crystals. The distance between the midpoints of the voltage contacts was 2.70$\pm$0.05 mm and the width of both voltage contacts together was 0.39$\pm$0.05 mm. If the distance between the outside and inside ends of contacts is used, then the value of resistivity will be 4$\%$ smaller and 11 $\%$ larger, respectively, of that calculated with the criterion described above. The cross-section area of the sample was 0.17$\times$0.09 mm$^2$ (W$\times$H).

The heat capacity was measured with the help of a relaxation technique. The measured background heat capacity, that includes sample platform and grease, for all necessary $(H,T)$ values was accounted for in the final results. The heat capacity of LaZn$_{11}$ was measured in the same temperature range and was used to estimate a non-magnetic contribution to the heat capacity of CeZn$_{11}$.

A four-wire geometry, $\it ac$ technique, was used to collect the Hall resistivity data. The polarity of the magnetic field was switched to remove any magnetoresitive components due to the misalignment of the voltage contacts. The current contacts were placed on two opposite side faces of the plate-shape crystals. The voltage contacts were placed on the other two remaining side faces of the crystals. All four contacts were made with the Pt wires and EpoTek H20E silver epoxy. 

The thermoelectric power (TEP) measurements were performed by a {\it dc}, alternating heating (two heaters and two thermometers) technique \cite{Mun2010b} using a Quantum Design PPMS to provide the temperature environment between 2 K and 300 K. The samples were mounted directly on the gold plated surface of the SD package of the Cernox thermometers using  Du-Pont 4929N silver paste to ensure thermal and electrical contact.

\section{Results and Analysis}
\subsection{Basic physical properties}

The powder x-ray diffraction pattern collected on the ground single crystals of CeZn$_{11}$ is shown in Fig. 1. X-ray diffraction measurements confirmed the crystal structure of CeZn$_{11}$. The main phase is CeZn$_{11}$ but small traces of Zn, from residual flux, can be detected as well. The lattice parameters obtained from the LeBail fit are a=b=10.67$\pm$0.02 {\AA} and c=6.87$\pm$0.01{\AA} and are consistent with the reported unit cell \cite{Zelinska2004}.

\begin{figure}[t]
\centering
\includegraphics[width=1\linewidth]{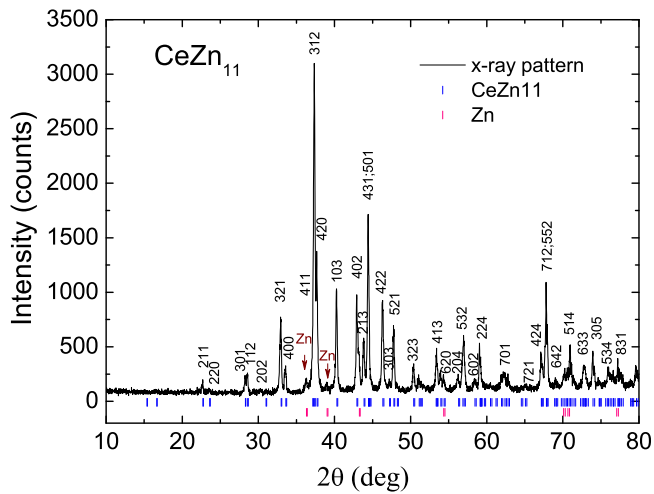}
\caption{\footnotesize (Color online) Powder X-ray diffraction pattern of finely ground CeZn$_{11}$ single crystals. Peaks are indexed to a tetragonal structure with a=10.67$\pm$0.02 {\AA} and c=6.87$\pm$0.01 {\AA}. A few, low intensity peaks can be assigned to the residual Zn flux.}
\end{figure}
\begin{figure}[t]
\centering
\includegraphics[width=1\linewidth]{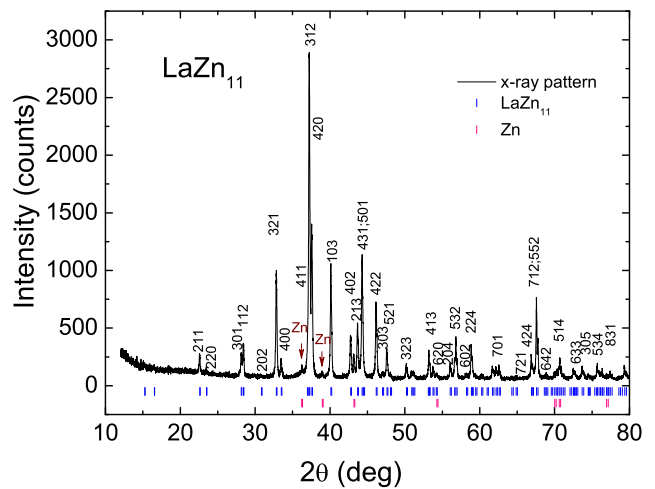}
\caption{\footnotesize (Color online) Powder X-ray pattern of fnely ground LaZn$_{11}$ single crystals. The peaks that belong to LaZn$_{11}$ are indexed to a tetragonal unit cell with a=10.69$\pm$0.02 {\AA} and c=6.89$\pm$0.01 {\AA}. The relatively few, low intensity peaks that can be associated to Zn are not indexed, however, the markers of the peaks positions are given. }
\end{figure}

\begin{figure}[tb]
\centering
\includegraphics[width=1\linewidth]{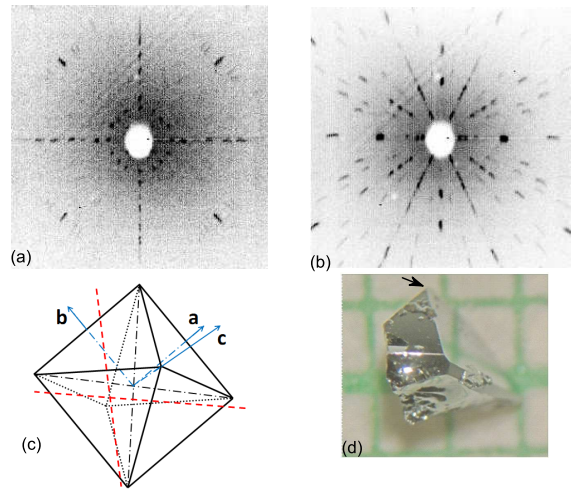}
\caption{\footnotesize (Color online) (a) X-ray Laue back scattering pattern showing a four fold rotation symmetry of the $\langle001\rangle$ direction, (b) X-ray Laue back scattering pattern showing a two fold rotation symmetry of the $\langle110\rangle$ direction, (c) a sketch of the sample with the main crystallographic directions and (d) the picture of the sample with the mirrored triangular surface being (101) plane. The apex that points out of the page is along the [001] direction, and the apex that is in the plane of the figure and points to top of the picture (marked with an arrow) is along the [110] direction. Dashed red (color online) lines in Fig. 3(c) indicate the boundaries of the sample shown in Fig. 3(d).}
\end{figure}

Figure 2 shows the powder x-ray diffraction pattern of powdered single crystals of LaZn$_{11}$. The lattice parameters for LaZn$_{11}$ obtained from the LeBail fit of the x-ray pattern are: a=b=10.69$\pm$0.02 {\AA} and c=6.89$\pm$0.01 {\AA} and are consistent with those published in the literature \cite{Iandelli1967}. 

Due to the samples' ambiguous morphology, slightly distorted octahedra, it is hard to visually identify the orientation of the facets. In order to determine the main crystallographic directions, Laue-back-reflection patterns were taken. Figures 3(a) and 3(b) show x-ray Laue back scattering patterns with a four fold rotation symmetry of the $\langle001\rangle$ direction and a two fold rotation symmetry of the $\langle110\rangle$ direction respectively. Figure 3(c) shows a sketch of the sample with the main crystallographic directions. Fig. 3(d) shows the picture of a sample (trancated by growth and separation) with the mirrored triangular surface being (101) plane. The apex that points out of the page is along the [001] direction, and the apex that is in the plane of the figure and points to top of the picture (marked with an arrow) is along the [110] direction. 

\begin{figure}[tb]
\centering
\includegraphics[width=1\linewidth]{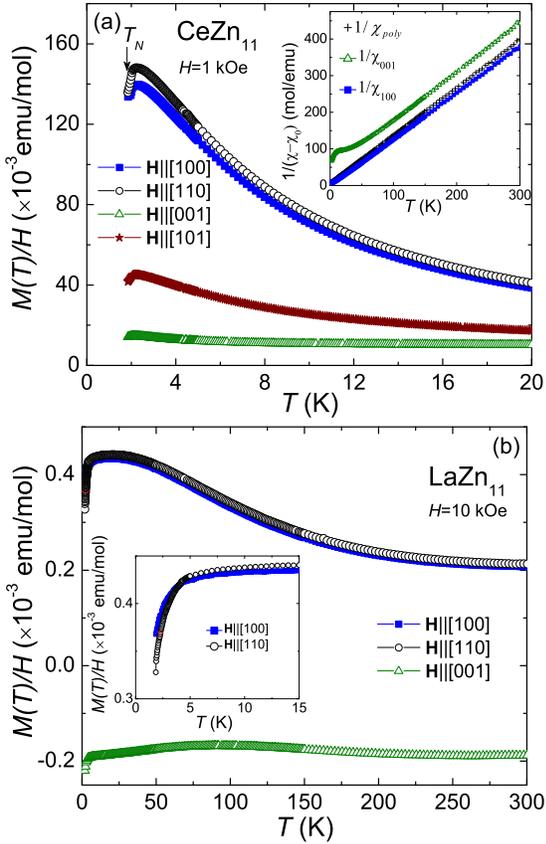}
\caption{\footnotesize (Color online) (a) Temperature-dependent magnetic susceptibility, $M(T)/H$, of CeZn$_{11}$, with the magnetic field applied along the [100], [110], [001], and [101] directions. The inset displays temperature-dependent, inverse magnetic susceptibilities for the field applied along the [100], [001] and of polycrystalline average $\chi_{poly}$=(2$\chi _{100}$+$\chi _{001}$)/3 .(b) Temperature-dependent magnetic susceptibility, $M(T)/H$, of LaZn$_{11}$, with the magnetic field applied along the [100], [110] and [001] directions. The inset shows the enlarged low temperature part of the $M(T)/H$. We would like to draw the reader's attention to the difference in the vertical scales of Figs. 4(a) and 4(b).}
\end{figure}

\begin{table}[t]
\caption{\label{} Results of the modified Curie-Weiss law fit of the magnetic susceptibility.}
\begin{ruledtabular}
\begin{tabular}{ccccc}
 $\chi$ & $\chi_0$ (emu/mol)& $\theta$(K) & $\mu_{eff}(\mu_B)$  \\ \hline
$\chi_{100}$ & (-7.8$\pm$0.1)$\times 10^{-4}$&1.2$\pm$0.5&\\
$\chi_{110}$ & (-7.8$\pm$0.1)$\times 10^{-4}$&-1.6$\pm$0.1&\\
$\chi_{001}$ & (-1.9$\pm$0.1)$\times 10^{-4}$&-31$\pm$1&\\
$\chi_{poly}$ & (-5.6$\pm$0.1)$\times 10^{-4}$&-5.6$\pm$1.0&2.48$\pm$0.01\\
\end{tabular}
\end{ruledtabular}
\end{table}

\begin{figure}[tb]
\centering
\includegraphics[width=1\linewidth]{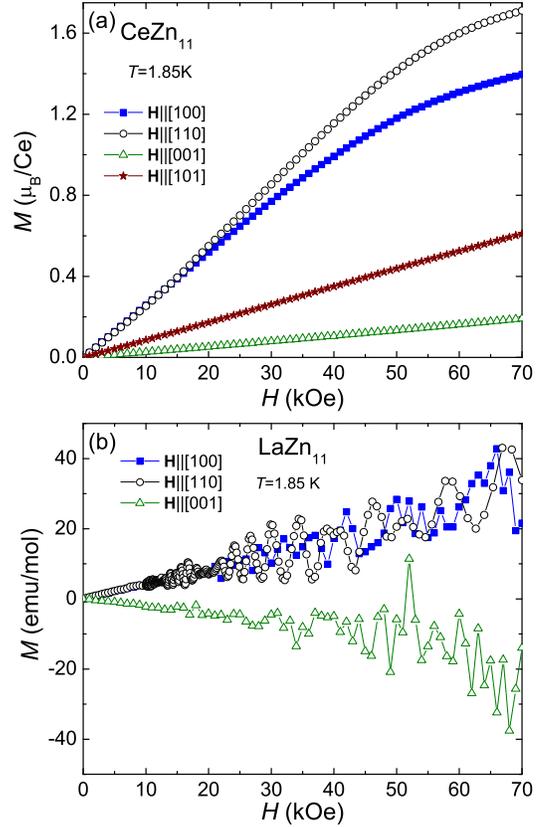}
\caption{\footnotesize (Color online) (a) Magnetization isotherms, $M(H)$, of CeZn$_{11}$ at 1.85 K for the magnetic field applied along the [100], [110], [001], and [101] directions. (b) Magnetization isotherms, $M(H)$, of LaZn$_{11}$ at 1.85 K for the magnetic field applied along the [100], [110] and [001] directions.}
\end{figure}

The anisotropic temperature-dependent magnetization divided by the applied field data, $M(T)/H$, of CeZn$_{11}$ are shown in Fig. 4(a), where the magnetic field was applied along the [100], [110], [001], and [101] directions. An arrow marks the value of $T_N$=2.00$\pm$0.03 K that was obtained from the maximum in d($\chi T$)/d$T$ \cite{Fisher1962} ($\chi$=$M(T)/H$ at small fields for which $M(H)$ at constant temperature is linear). The magnetic susceptibilities for the field applied along the [100] and [110] directions are essentially the same and are about 10 times larger than the magnetic susceptibility for the field applied along the [001] direction. As would be expected, when the field is applied along the [101] direction, $M(T)/H$ falls between the data for $\bf{H}\|$[100] and $\bf{H}\|$[001]. The inset of Fig. 4(a) displays temperature-dependent inverse magnetic susceptibilities, $H/M(T)$, for the field applied along the [100], [001] and of polycrystalline average taken as $\chi_{poly}$=(2$\chi _{100}$+$\chi _{001}$)/3. The modified Curie-Weiss law fit of the magnetic susceptibility in the form $\chi$ =$\chi _0$+$C/(T-\theta$) of the polycrystalline average above 50 K results in $\theta _p$ $\simeq$-5.6$\pm$1.0 K and $\mu _{eff}$ $\simeq$2.48$\pm$0.01 $\mu _B$/Ce$^{3+}$ which is close to the Ce$^{3+}$ free ion value of 2.54 $\mu _B$. The results of the modified Curie-Weiss law fit are listed in Table 1. When $\chi_0$ of CeZn$_{11}$ is compared with the high temperature $\chi(T)$ of LaZn$_{11}$, there is a good agreement for $\bf{H}\|$[001] and poor agreement for in-plane measurements which is most likely due to small changes in terms with differing signs (Pauli versus Landau and Larmor diamagnetism).

The $M(T)/H$ data for LaZn$_{11}$ are shown in Fig. 4(b), where the magnetic field was applied along the [100], [110] and [001] directions. $M(T)/H$ for $\bf{H}\|$[100] and [110] are positive whereas $M(T)/H$ for $\bf{H}\|$[001] is negative, reflecting anisotropy of the Pauli and Landau terms relative to the Larmor diamagnetic susceptibility. The apparent drop in $M(T)/H$ of LaZn$_{11}$ below $\sim$10 K is due to formation/filling of Landau levels associated with quantum oscillations shown in detail in Fig. 5(b). 

Magnetic isotherms, $M(H)$, of CeZn$_{11}$, shown in Fig. 5(a), were taken at 1.85 K for the magnetic field applied  along the same orientations as the $M(T)/H$ in Fig. 4(a). The magnetization does not saturate in any direction for our highest field of 70 kOe. The magnetic moment is highly anisotropic and reaches the highest value of 1.7 $\mu _B$/Ce$^{3+}$ at 70 kOe for the magnetic field applied along the [110] direction, which is below the theoretical saturated value of 2.1 $\mu _B$ for the free Ce$^{3+}$ ion. In addition, $M(H)$ for the magnetic field along the [110] direction, shows a broad feature centered at about 18 kOe, which, as will be shown in detail below, is associated with the suppression of $T_N$ to below 1.85 K at this field.

Magnetic isotherms, $M(H)$, of LaZn$_{11}$, shown in Fig. 5(b), were taken at 1.85 K for the magnetic field applied  along the same orientations as the $M(T)/H$ in Fig. 4(b). Consistent with the $M(T)$ data, $M(H)$ for $\bf{H}\|$[001] is negative whereas $M(H)$ for $\bf{H}\|$[100] and $\bf{H}\|$[110] are positive. The de Haas-van Alphen oscillations are clearly seen setting in near $\sim$10 kOe in the $M(H)$ for all field orientations. Analysis of the quantum oscillations is given in the Appendix below. Change in the amplitude of the quantum oscillations with temperature is the origin of the drop in the $M(T)/H$ seen in Fig. 4(b). As shown in the inset of Fig. 4(b), for $H$=10 kOe the Landau levels start to develop and fill below $T\sim$ 10 K.

\begin{figure}[tb]
\centering
\includegraphics[width=1\linewidth]{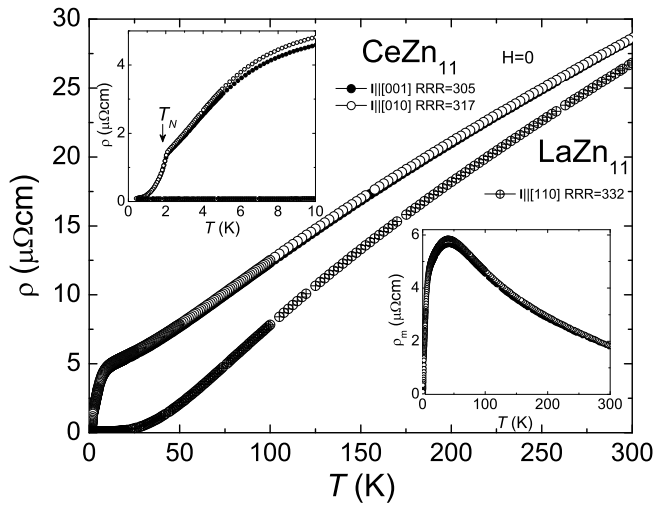}
\caption{\footnotesize Zero-field, temperature-dependent resistivity, $\rho(T)$, of CeZn$_{11}$ and LaZn$_{11}$. The inset on the upper left shows an enlarged, low-temperature, part of the resistivity with the AFM transition, $T_N$, marked with an arrow. The inset on the lower right shows magnetic part of the resistivity, $\rho_m$=$\rho$(CeZn$_{11}$)-$\rho$(LaZn$_{11})$.}
\end{figure}

\begin{figure}[tb]
\centering
\includegraphics[width=1\linewidth]{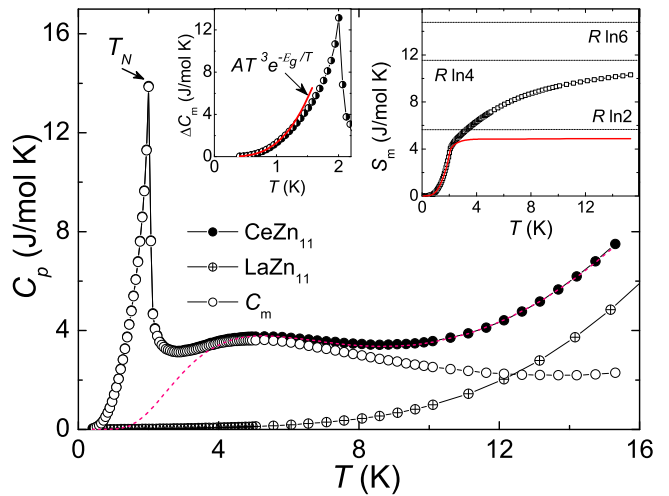}
\caption{\footnotesize Heat capacity, $C_p(T)$, of CeZn$_{11}$ ($\bullet$  -total), $\oplus$ - heat capacity of LaZn$_{11}$ (electronic and lattice), and $\circ$ - magnetic ($C_m$=$C_p$(CeZn$_{11}$)-$C_p$(LaZn$_{11}$)). Dashed curve: modeled Schottky anomaly assuming that the energies of the 1st and 2nd excited states are $\Delta_1$=12.2 K and $\Delta_2$=65 K added to heat capacity data of LaZn$_{11}$. The inset on the upper left shows $\Delta C_m=C_p(CeZn_{11})-C_p(LaZn_{11})-C_{Sch}$, where $C_{Sch}$ is the contribution of the modeled Schottky anomalies, as a function of temperature. Red line is the low-temperature fit of the data with the function, $AT^3e^{-E_g/T}$, that is expected for the magnon specific heat of  antiferromagnet with the energy gap, $E_g$, in the magnon dispersion relation \cite{Gopal1966,Tari2003}. $A$ is the coefficient. The inset on the upper right shows magnetic entropy,  $S_m$, calculated by integrating $C_m/T$ ($\circ$) and $\Delta C_m/T$ (line).}
\end{figure}

The zero-field temperature-dependent resistivity, $\rho(T)$, of CeZn$_{11}$, presented in Fig. 6, for current flow along the [010] and [001] directions, shows a broad shoulder, characteristic of that of Kondo compounds, at around 10 K followed by a sharp change of the slope and a kink corresponding to the AFM transition at 1.96$\pm$0.05 K estimated from the maximum in d$\rho$/d$T$ \cite{Fisher1968}. The broad shoulder may also have some contribution associated with the relatively small value of the CEF splitting, which stems from the very symmetric, local, environment of the Ce ion (in a shell of 22 Zn atoms) \cite{Zelinska2004}. $\rho(T)$ for both current directions is similar, although, as was mentioned above, the virtually equal resistivity values for both current orientations is most likely a coincidence and merely means that $\rho(T)$ has relatively low anisotropy. The residual resistivity ratios (RRR) of the two samples are 305 for $\bf{I}\|$[001] and 317 for $\bf{I}\|$[100]. The inset on the upper left of Fig. 6 shows low temperature part of resistivity of CeZn$_{11}$ with the AFM transition, $T_N$, marked with an arrow.

The zero-field temperature-dependent resistivity, $\rho(T)$, of LaZn$_{11}$, the non-magnetic, isostructural variant of CeZn$_{11}$, is also shown in Fig. 6 for comparison and gives an estimate of the non-magnetic contribution to the resistivity of CeZn$_{11}$. With the assumption that resistivity of LaZn$_{11}$ is isotropic, we calculated the magnetic resistivity, $\rho_m$=$\rho$(CeZn$_{11}$)-$\rho$(LaZn$_{11})$, which is given in the lower right inset of Fig. 6. $\rho_m$ shows a broad peak near 40 K which maybe associated with a combination of possible Kondo physics with the certain thermal depopulation of the exited CEF level as the temperature is decreased. 

Heat capacity data, $C_p(T)$, for CeZn$_{11}$, Fig. 7, show a clear, sharp, $\lambda$ anomaly at 2.00$\pm$0.06 K, which is consistent with magnetization and resistivity measurements, all of which are consistent with a second order transition from the paramagnetic to an AFM ordered state. A broad peak at around 5 K corresponds to the Schottky anomaly arising from the CEF splitting of the Hund's rule ground state multiplet. The sum of the electronic and lattice contribution to the specific heat of CeZn$_{11}$ may be approximated by the specific heat of LaZn$_{11}$ which is also shown in Fig. 7. The electronic specific heat coefficient ($\gamma$) and the Debye temperature ($\Theta_D$) estimated from the relation $C_p/T=\gamma + \beta T^2$ for LaZn$_{11}$ are 10.2 mJ/(mol K$^2$) (or 0.85 mJ/(mol K$^2$ atom)) and 353 K ($\beta$=0.53 mJ/(mol K$^4$)) respectively and are similar to the ones reported in Ref. \cite{Nakazawa1993}. Because of the AFM order at 2 K and the higher temperature broad peak associated with CEF effect, $\gamma$ and $\Theta_D$ for CeZn$_{11}$ cannot be estimated from the relation $C_p/T=\gamma + \beta T^2$. As a matter of fact, the fit of the $C_p/T$ vs $T^2$ gave $\gamma$ values in the range 100-200 mJ/(mol K$^2$) depending on the temperature range chosen. $C_p/T$ of CeZn$_{11}$ has value of 166 mJ/(mol K$^2$) at 0.4 K and is still significantly larger than the value of LaZn$_{11}$ (4.8 mJ/(mol K$^2$)) at the same temperature.

The magnetic contribution to the specific heat of CeZn$_{11}$, calculated as $C_m$=$C_p$(CeZn$_{11}$)-$C_p$(LaZn$_{11}$), allows for the inference of the change in the entropy by integrating $C_m/T$ with respect to $T$, shown in the inset of Fig. 7. The very slight mass-correction to the LaZn$_{11}$ data \cite{Bouvier1991} was not done since the mass-correction factor $\Theta_D$(CeZn$_{11}$)/$\Theta_D$(LaZn$_{11}$) is 0.9990.  The entropy removed up to $T_N$ is 0.64$R$ ln2 (inset of Fig. 7) and is consistent with the ordered state emerging from a CEF ground state doublet with the first excited state doublet located nearby. According to Ref. \cite{Gopal1966}, for two-level system that has equal degeneracy, the maximum of the Schottky anomaly occurs at $T_m$=0.42$\delta$, $\delta$ is the energy separation between two levels, and $C_{Sch}(max)$=3.64 J/(mol K). For CeZn$_{11}$, $C_{Sch}(max)$=3.62 J/(mol K) at $T_m\sim$5.31 K then $\delta\sim$12.6 K. An alternate way to estimate the CEF splitting is to fit the $C_m$ data to the Schottky anomalies assuming that the $J$=5/2 multiplet is split into three doublets by tetragonal point symmetry. But taking into account that the peak associated with the AFM transition is very close to the Schottky anomaly, instead of fitting $C_m$ data, we modelled the Schottky anomalies assuming that the energies of the 1st and 2nd doubly degenerate excited states are $\Delta_1$ and $\Delta_2$ and added the modelled Schottky anomalies to the heat capacity data of LaZn$_{11}$. The result is illustrated by a dashed curve in Fig. 7. By adjusting the values of $\Delta_1$ and $\Delta_2$ we obtained the best agreement with the experimental data for $\Delta_1$=12.2 K and $\Delta_2$=65 K, which are slightly higher than the ones reported in Ref. \cite{Nakazawa1993}. If we perform the fit of $C_{mag}(T)$ as $\gamma' T$+$C_{Sch}$, where $C_{Sch}$ is the contribution of the modeled Schottky anomalies, then  $\gamma'$=24.3 mJ/(mol K$^2$) and $\Delta_1$=11.3 K and $\Delta_2$=71.4 K. In this case $\gamma(total)$=$\gamma$(LaZn$_{11})+\gamma'$ for CeZn$_{11}$.
To find out the functional dependence of the magnetic specific heat at low temperatures, below the AFM transition, we fitted the data below 1 K of $\Delta C_m=C_p(CeZn_{11})-C_p(LaZn_{11})-C_{Sch}$, where $C_{Sch}$ is the contribution of the modeled Schottky anomalies, with the function, $AT^3e^{-E_g/T}$, that is expected for the magnon specific heat of antiferromagnet with the energy gap, $E_g$, in the magnon dispersion relation \cite{Gopal1966,Tari2003}. $A$ is the coefficient. The result of the fit is illustrated as a solid line in the inset on the upper-left of Fig. 7. From the fit we obtained $E_g$=0.66 K. An integral of $\Delta C_m/T$, a magnetic entropy, is shown as a solid line in the inset in Fig. 7. The magnetic entropy is $\sim$0.85$R$ ln2 which is consistent with the ordered state emerging from the CEF state doublet.

\begin{figure}[tb]
\centering
\includegraphics[width=1\linewidth]{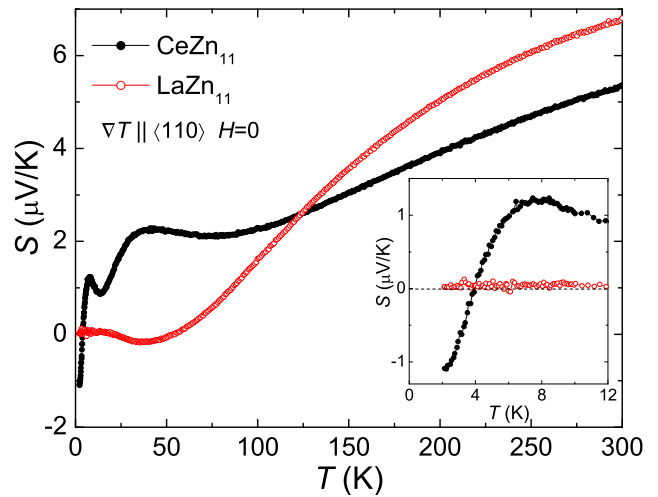}
\caption{\footnotesize Zero-field, temperature-dependent, thermoelectric power, $\it S(T)$, of CeZn$_{11}$ and LaZn$_{11}$. The inset: expanded, low-temperature portion of the plot. }
\end{figure} 

\begin{figure}[tb]
\centering
\includegraphics[width=1\linewidth]{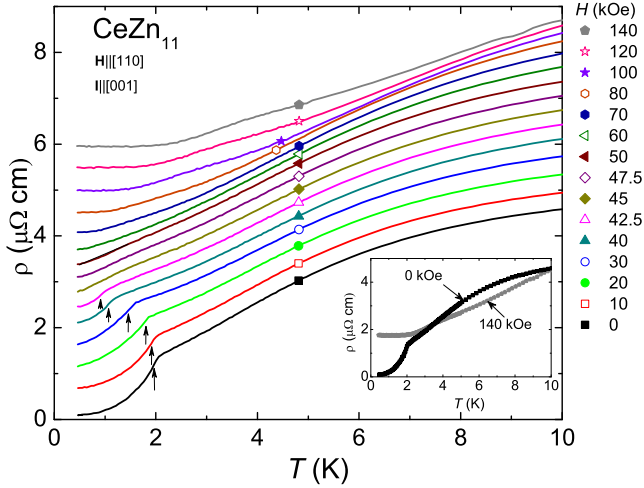}
\caption{\footnotesize (Color online) Low-temperature parts of $\rho(T)$ curves for CeZn$_{11}$ taken at different applied fields for $\bf{H}\|$[110]. Subsequent data sets are shifted upward from each other by 0.3 $\mu \Omega$ cm for clarity. For $H\leq$ 42.5 kOe, the arrows denote the transition temperature, $T_N$, inferred from d$\rho$/d$T$. The inset shows the resistivity curves for $H$=0 and $H$=140 kOe without any offset.}
\end{figure}

\begin{figure}[tb]
\centering
\includegraphics[width=1\linewidth]{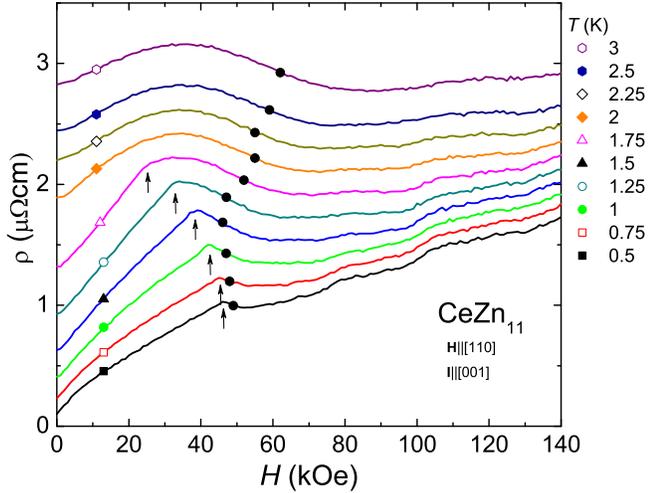}
\caption{\footnotesize (Color online) $\rho(H)$, ($\bf{H}\|$[110]), isotherms for CeZn$_{11}$. Subsequent data sets are shifted upward by 0.1 $\mu \Omega$ cm from each other. Arrows denote the field of magnetic ordering inferred from d$\rho$/d$H$. Solid dots represent the position of the broad minimum inferred from d$\rho$/d$H$.}
\end{figure}

\begin{figure}[t]
\centering
\includegraphics[width=1\linewidth]{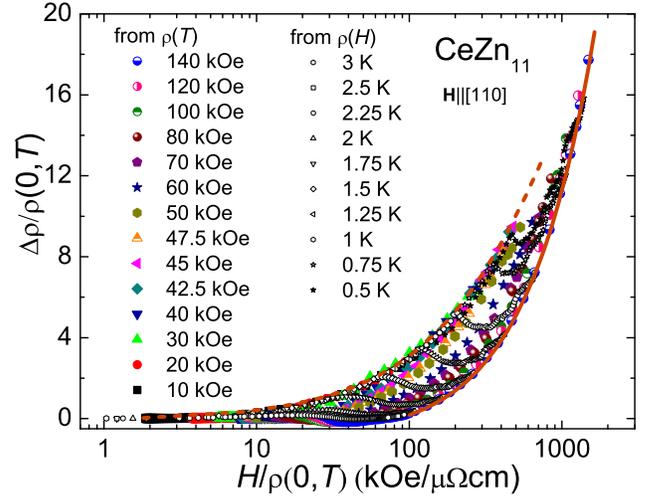}
\caption{\footnotesize (Color online) Kohler plot (showing every other data point), $\Delta\rho/\rho(0,T)$, where $\Delta\rho=\rho(H,T)-\rho(0,T)$, for CeZn$_{11}$ for the field applied along the [110] direction. The dashed and solid lines denote two manifolds that the data appear to follow.}
\end{figure}

The zero-field, temperature-dependent, thermoelectric power (TEP), $\it S(T)$, of CeZn$_{11}$ is plotted in Fig. 8. Since the base temperature for the TEP measurements is 2.1 K, the AFM transition for CeZn$_{11}$ was not observed in the TEP measurements. The value of the TEP is positive above 4 K signifying that hole-type carriers dominate the thermoelectric transport in this material. The temperature dependence of the TEP for CeZn$_{11}$ is reminiscent of $S(T)$ of the noble metals \cite{Blatt1976} or Zn, although it does have a complex behaviour rather similar to $S(T)$ of the Ce-based heavy fermion compounds which usually displays one or several peaks \cite{Jaccard1990,Franz1978,Jaccard1982}. The absolute value of the TEP is not anomalously large (TEP of LaZn$_{11}$ being larger for $T>$130 K) and the temperature dependence of the TEP is almost linear above 125 K. Two positive maxima, at $\sim$8 K and $\sim$40 K and a positive minimum at $\sim$13.6 K are observed in the $\it S(T)$ at temperatures below 60 K. Since $S$ must vanish as $T$ tends to zero, our data, inset of Fig. 8, suggest the occurrence of at least one more negative extremum in the TEP data as the temperature is lowered below 2.1 K. The position of the peak at $\sim$40 K is very similar to the one found in the TEP of Zn (with the temperature gradient perpendicular to the hexagonal axis \cite{Rowe1970}) that has been attributed to phonon drag. 

The zero-field temperature-dependent thermoelectric power, $\it S(T)$, of LaZn$_{11}$ is also plotted in Fig. 8. As can be seen from the inset of Fig. 8, $S(T)$ of LaZn$_{11}$ tends to zero as temperature is lowered. The broad negative minimum at $\sim$37 K is probably due to the phonon drag contribution to the TEP \cite{Blatt1976,Elliott1972}. 

Based on our modelled Schottky anomalies, the lower temperature of CEF splitting, $\Delta_1$ is likely to be the origin of the broad shoulder observed in $\rho(T)$ at around 10 K and also of positive maximum in the TEP at around 8 K. On the other hand, the origin of the positive maximum in the TEP around 40 K, the position of which is weakly affected by the applied magnetic field (shown below in Fig. 16) and which is at similar temperature as the broad peak in $\rho_m(T)$, may be ascribed to an interplay of Kondo, crystal-field and phonon drag effects. If we assume that the $\theta_D$ of CeZn$_{11}$ is very close to that of LaZn$_{11}$, then the peak due to the phonon-drag contribution to the TEP data should be expected at 0.1-0.3$\theta_D$ \cite{Blatt1976, Elliott1972} which translates into 35-106 K temperature range. The negative minimum at $\sim$37 K in $S(T)$ of LaZn$_{11}$ and positive maximum at $\sim$40 K in $S(T)$ of CeZn$_{11}$ are both in this temperature range.

\begin{figure*}[tb]
\centering
\includegraphics[width=0.75\linewidth]{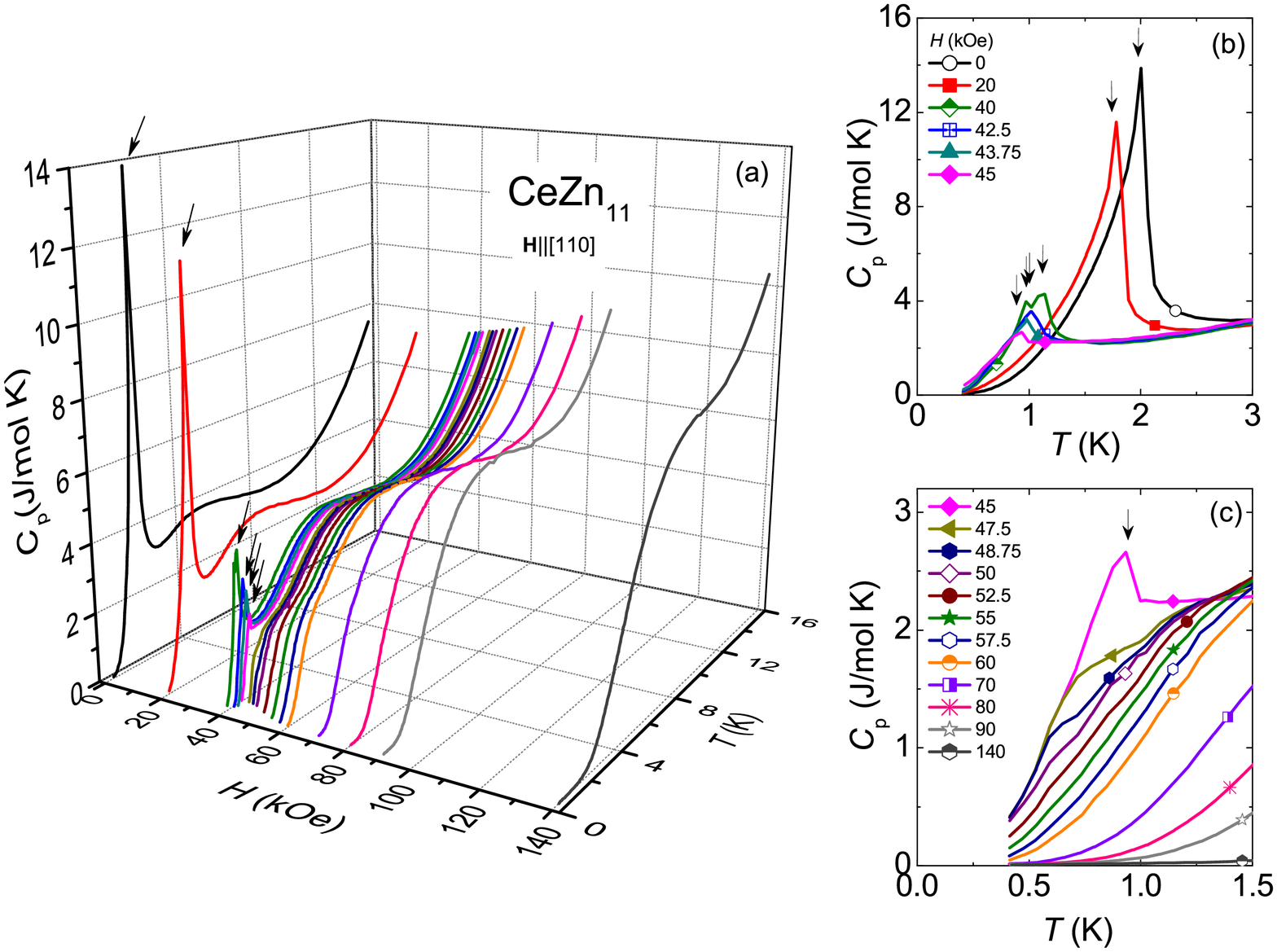}
\caption{\footnotesize(Color online) (a) Heat capacity data for CeZn$_{11}$ taken in the applied magnetic field, $\bf{H}\|$[110], (b) and (c) expanded, low-temperature part of heat capacity curves. The arrows indicate peaks associated with the magnetic ordering. 
}
\end{figure*}

\begin{figure}[tb]
\centering
\includegraphics[width=1\linewidth]{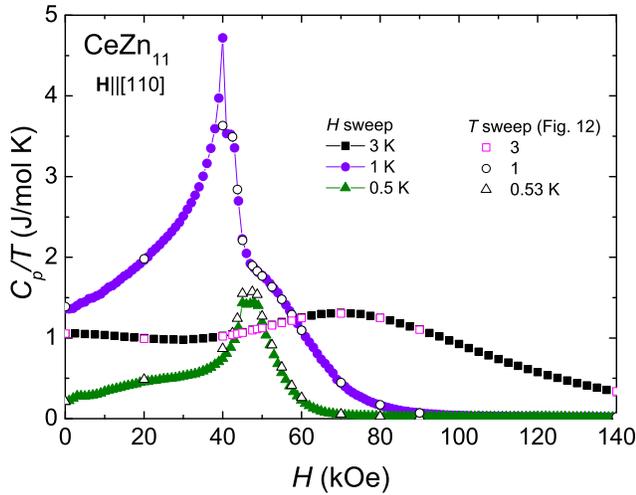}
\caption{\footnotesize (Color online) $C_p/T$ of CeZn$_{11}$ as a function of magnetic field measured at constant temperatures of 0.5, 1, and 3 K (solid symbol data) with the magnetic field along the [110] direction. Open symbol data are taken from $C_p(T)|_{H=const}$ data in Fig. 12. }
\end{figure}

\begin{figure}[tb]
\centering
\includegraphics[width=1\linewidth]{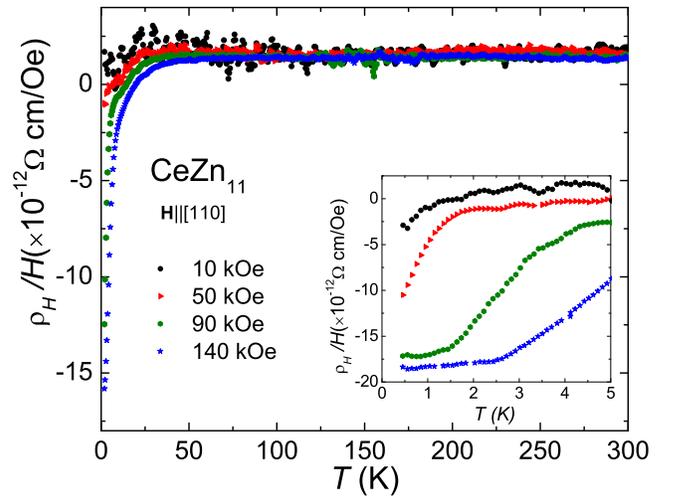}
\caption{\footnotesize (Color online) Hall coefficient, $R_H=\rho_H$/$H$, versus temperature for CeZn$_{11}$ at $H$=10, 50, 90, and 140 kOe. The inset: low-temperature part of the Hall coefficient data. The data sets were smoothed with the adjacent-point-averaging method with a 5 points of window. The current was applied along the [1$\bar{1}$0] direction and the magnetic field was applied along the [110] direction.}
\end{figure}

\begin{figure}[tb]
\centering
\includegraphics[width=1\linewidth]{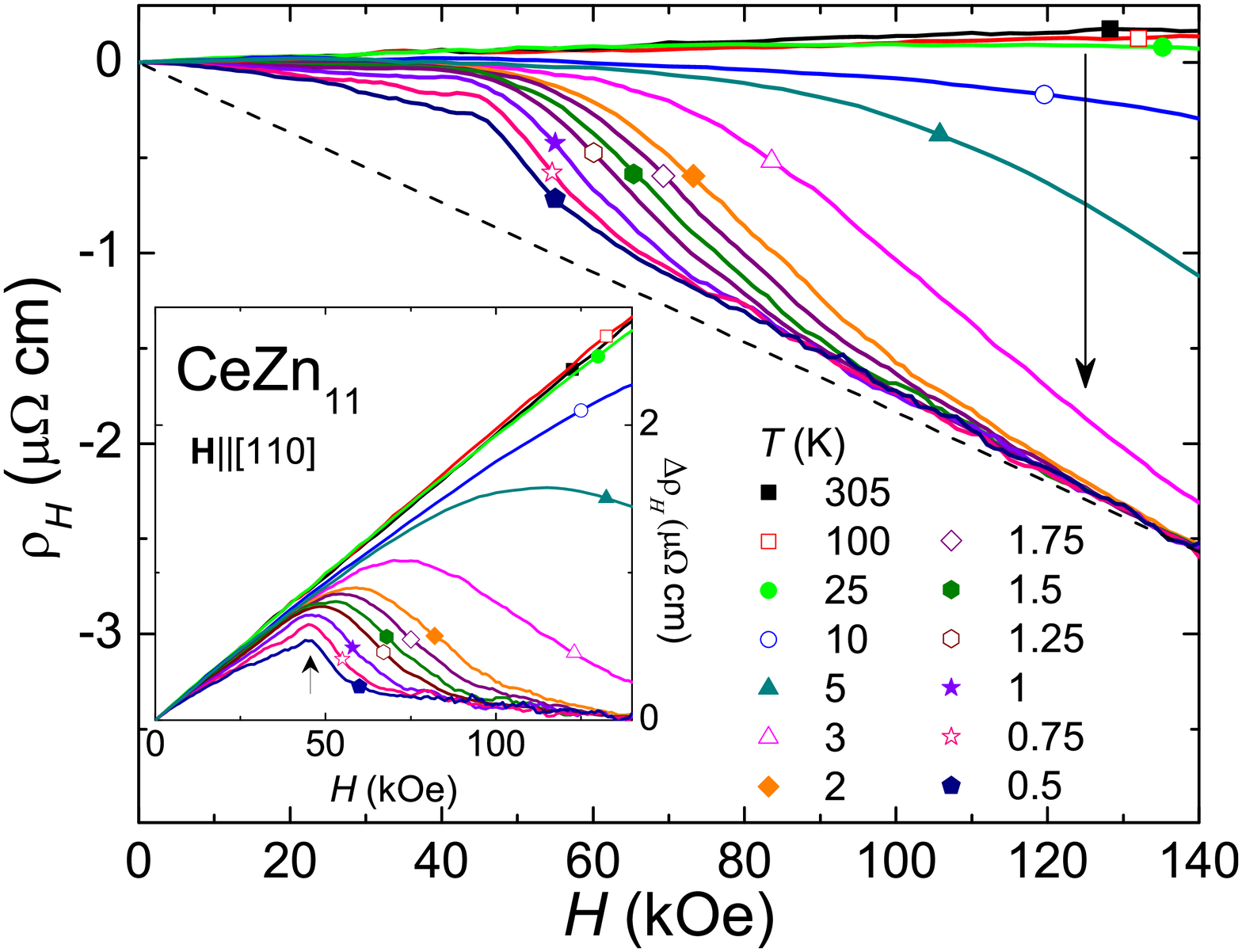}
\caption{\footnotesize (Color online) Hall resistivity, $\rho_H(H)$, of CeZn$_{11}$ measured at different constant temperatures. The data sets were smoothed with the adjacent-point-averaging method with a 5 points of window. The inset shows the data after the linear background, represented by a dashed line on the main graph, is subtracted. The arrow indicates the field of the maximum which corresponds to a transition from ordered to non-ordered state at the lowest temperature. The current was applied along the [1$\bar{1}$0] direction and the magnetic field was applied along the [110] direction. 
}
\end{figure}

\subsection{Measurements in applied magnetic field ($\bf{H}\|$[110])}
Given its relatively low $T_N$ value, coupled with our ability to grow very high quality, large RRR single crystals, we decided to determine the $T-H$ phase diagram for CeZn$_{11}$ and look for possible quantum critical effects.
Since Fig. 5(a) shows that the magnetization for the magnetic field applied along the [110] direction is larger than that for the field applied along any other measured direction, indicating that [110] is likely to be an easy axis, we will focus our attention on the field dependence of measurements with $\bf{H}\|$[110]. Figure 9 shows low-temperature $\rho(T)$ curves for CeZn$_{11}$ taken at different applied fields for $\bf{H}\|$[110] with each subsequent data sets shifted upward by 0.3 $\mu \Omega$ cm for clarity. The arrows denote the transition temperature, $T_N$, inferred from the maximum in d$\rho$/d$T$ \cite{Fisher1968}. As the magnetic field is increased, the kink associated with the AFM transition temperature moves to the lower temperatures and is suppressed below our base temperature of 0.46 K by $H$=45 kOe. For the applied field of 45 kOe, the low temperature $\rho(T)$ functional dependence appears to be essentially linear, $\rho(T)\sim T$, from base temperature of 0.46 K to $\sim$2 K. Finally, for the applied field larger than 60 kOe, at low temperature an upturn in the $\rho(T)$ appears. 

$\rho(H)$ isotherms for CeZn$_{11}$ are shown in Fig. 10. The arrows denote the transition, from the low field ordered state to a higher field most likely saturated paramagnetic state, inferred from d$\rho$/d$H$. Another feature in the field dependent resistivity appears above 45 kOe and can be seen more clearly in the d$\rho$/d$H$, as will be shown and discussed  below, and is marked with solid dots in Fig. 10. This feature moves to the higher fields as the temperature is increased. Small amplitude Shubnikov-de Haas (SdH) oscillations can be seen for the applied magnetic fields larger than 70 kOe in the lower temperature data.

Figure 11 shows the Kohler plot for CeZn$_{11}$ for the field applied along the [110] direction. It is clear from the Kohler plot that the magnetoresistance of CeZn$_{11}$ fails to show a simple quadratic dependence at all applied fields measured and it appears to have two manifolds: one before ([$H/\rho(0,T)]^{0.7}$) and one after ([$H/\rho(0,T)]^{1}$) the AFM transition, they are represented by the dashed and solid lines respectively in Fig. 11.

The heat capacity measurements of CeZn$_{11}$ taken at different applied magnetic fields for the magnetic field applied along the [110] direction are plotted in Fig. 12. The arrows in Figs. 12(a), 12(b) and 12 (c), indicating the position of the peaks associated with the magnetic ordering, are consistent with the resistivity measurements described above. As the applied magnetic field is increased, the sharp peak corresponding to the AFM transition moves to lower temperatures and decreases in size. From $H$=47.5 kOe to $H$=52.5 kOe, the lower field peak in the specific heat evolves into a rather broad shoulder (see Fig. 12(c)). 

Figure 13 presents the heat capacity data of CeZn$_{11}$ as a function of magnetic field at constant temperatures of 0.5, 1, and 3 K. At 0.5 K, field-dependent specific heat appears to be almost constant over the 45-47 kOe field range which may be indicative of multiple phase transitions or a fact that the region of the $T-H$ phase diagram, shown below, that is centered at this range of magnetic field, is rich and complex. At 1 K, $C_p(H)$ manifests a sharp peak at 40 kOe, that corresponds to transition from ordered to non-ordered state, followed by two features, that might be related to the first order nature of the AFM transition, as the magnetic field is further increased. At 3 K, only a broad feature is observed at $\sim$70.5 kOe in $C_p(H)$. As could be seen from Fig. 13, the field sweep data (solid symbols) agree well with the data taken from temperature sweep in constant magnetic field (open symbols) that were shown in Fig. 12. 

To shed more light on the field induced states of CeZn$_{11}$ for $\bf{H}\|$[110], we further characterized the transport properties of CeZn$_{11}$ with the Hall resistivity, $\rho_H$, and thermoelectric power, $S$. 
Figure 14 shows the Hall coefficient, $R_H$=$\rho_H/H$, for CeZn$_{11}$. Above 50 K, $R_H$ is essentially temperature and field independent with the positive value indicating that hole-like carriers dominate in the electrical transport, which is consistent with what was observed in the zero-field TEP data in Fig. 8. Upon cooling below 1.8 K for $H$=10 kOe, $R_H$ changes sign from positive to negative. The temperature at which $R_H$ changes sign increases with the increase of the applied field, at $H$=140 kOe, $R_H$ changes sign at $\sim$20 K. The features seen at low temperatures (see inset in Fig. 14) in the data for $H$=50, 90 and 140 kOe possibly are related to a crossover from $\omega\tau\gg$1 to $\omega\tau\ll$1 limit ($\omega$ is the cyclotron frequency) which are correlated with the dHvA oscillations that were seen in the resistivity data in Fig. 10. \cite{Abrikosov1988} In the one-band approximation, the carrier density at 300 K is n$\simeq$4.4$\times$10$^{28}$ m$^{-3}$ ($R_H$=1.4$\times$ 10$^{-12}$ $\Omega$ cm/Oe) and is closer to the carrier density of silver $\simeq$5.86$\times$10$^{28}$ m$^{-3}$ rather than to that of zinc $\simeq$13.2$\times$10$^{28}$ m$^{-3}$.\cite{Ashcroft1976} Although it might be useful for some context, we would like to remark that the one-band and $m=m_e$ approximation is a gross oversimplification when applied in CeZn$_{11}$ case. 

Figures 15 shows $\rho_H(H)$ measured at several constant temperatures. For the temperatures higher than 25 K, $\rho_H$ is positive and almost linear. For the temperatures 10 K and below, $\rho_H$ not only becomes non-linear but also changes sign from positive to negative. As the temperature is decreased, the field at which $\rho_H$ changes sign moves toward lower applied magnetic fields. As will be discussed below, the sharp, low field feature seen in the lowest temperature data is associated with the transition from the low field, AFM state to a higher field state that is most likely saturated paramagnetic state. To see this feature clearer and to track its progression as we increased the temperature, we subtracted the linear background, represented by a dashed line in the main graph. The result is shown in the inset to Fig 15. The sharp peak, that corresponds to the AFM transition at two lowest temperatures, evolves into a broad peak positioned at a higher applied field as the temperature is increased. CEF effects are probably the origin of this broad feature. 

Figure 16 shows the semi-log plot of the temperature-dependent thermoelectric power, $\it S(T)$, of CeZn$_{11}$ taken at different applied magnetic fields. As the applied field is increased, the temperature of the peak positioned at $\sim$ 40 K first moves slightly up, and then moves more rapidly down and that of the lower peak at $\sim$ 8 K moves up and seems to merge with the lower temperature tail of the one at $\sim$ 40 K at 140 kOe (see inset to Fig. 16). Above $\sim$100 K, the TEP stays almost unchanged with the increase of the applied magnetic field and has almost linear temperature dependence. For $H$=0 TEP reverses the sign at 4 K: $S<$0 for $T<$4 K. Then, as the applied field is increased, the temperature at which the sign is changed moves to the higher temperatures. For $H\geq$120 kOe, the TEP changes sign twice. For example, for $H$=120kOe, $S>$0 for $T>$6.3 K and $T<$2.2 K. Within experimental temperature window accessible with our current measurements, it is not possible to determine exactly the magnetic field at which TEP starts to change sign second time, although, it seems that the required field is closer to $H$=120 kOe rather than to $H$=100 kOe.

The TEP as a function of the field, $S(H)$, taken at 2.3 K is plotted in Fig. 17 and has a complex behavior. Quantum oscillations are seen above $\sim$70 kOe in the $S(H)$. The amplitude of the quantum oscillations observed in the TEP is much larger than the ones observed in other measurements which is a unique aspect of the TEP because, rather than depend on the density of states at the Fermi level, it depends on the derivative of the density of states evaluated at the Fermi level, provided that the density of states is changing at the Fermi level. \cite{Fletcher1981,Mun2011} The fast Fourier transform (FFT) analysis of the quantum oscillations seen in the TEP measurements is given in the inset of Fig. 17 and shows strong peaks in the FFT spectrum at $\sim$2.0 MG ($\alpha$), $\sim$4.1 MG ($\beta$), $\sim$5.9 MG ($\gamma$), $\sim$10.2 ($\delta$), $\sim$12.4 MG ($\varepsilon$), and $\sim$14.4 ($\zeta$). To obtained the FFT spectrum, we took the last 1024 data points and subtracted the background that was fitted with the polynomial function. After that the data were plotted in 1/$H$ and interpolated so that the 1024 data points  were equally spaced. After that the FFT analysis was done.

\begin{figure}[tb]
\centering
\includegraphics[width=1\linewidth]{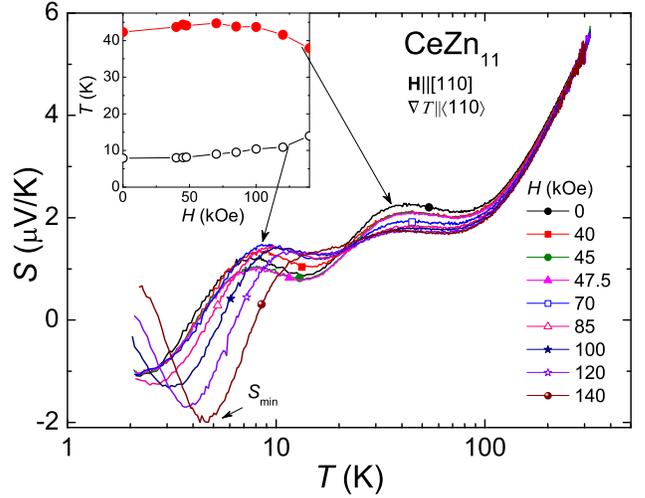}
\caption{\footnotesize (Color online) The semi-log plot of the temperature-dependent thermoelectric power, $\it S(T)$, of CeZn$_{11}$ taken at different applied magnetic field. The inset in the upper left corner illustrates evolution of two maxima as a function of applied field. }
\end{figure}

\begin{figure}[tb]
\centering
\includegraphics[width=1\linewidth]{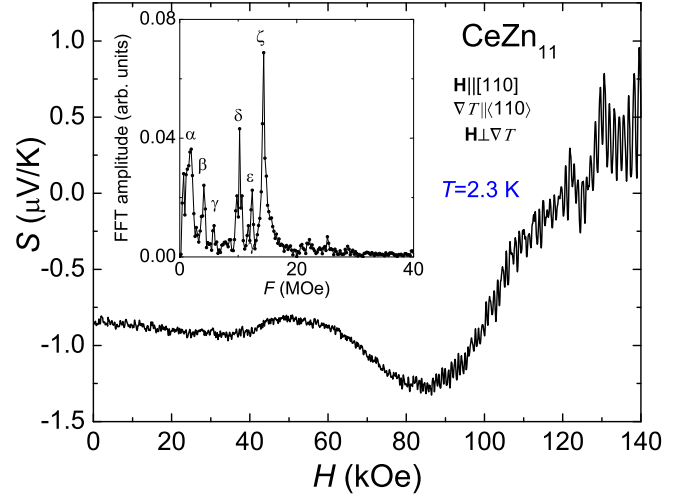}
\caption{\footnotesize (Color online) (a) Field-dependent thermoelectric power, $\it S(H)$, of CeZn$_{11}$ at 2.3 K. The inset: FFT spectrum of the oscillations obtained from TEP at 2.3 K}
\end{figure}

\begin{figure*}[tb]\centering
\includegraphics[width=1\linewidth]{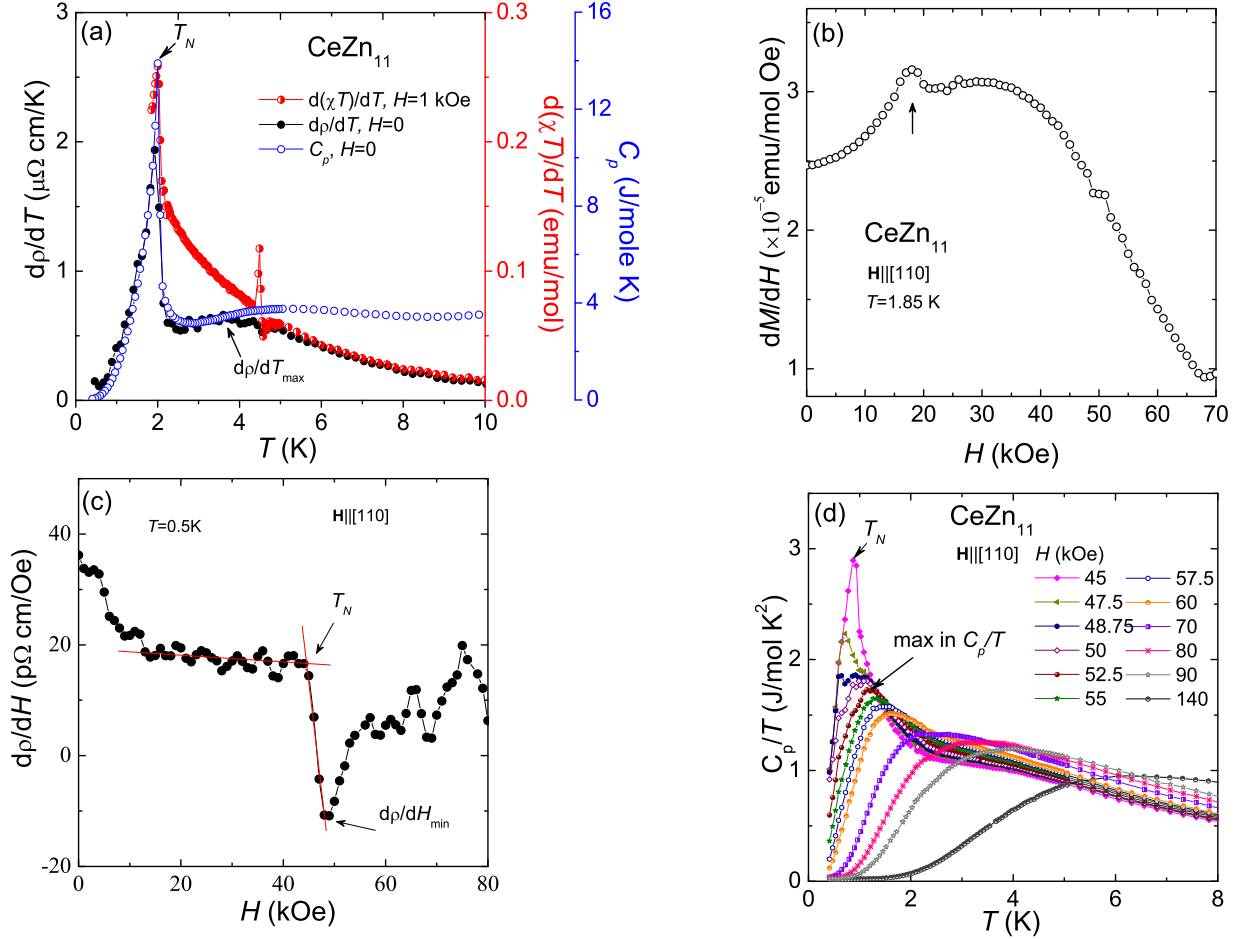}
\caption{\footnotesize (Color online) (a) d$\rho$/d$T$, d($\chi T)$/d$T$ and $C_p$ as a function of temperature. Arrows denote the AFM transition temperature and a broad shoulder in d$\rho$/d$T$. The $\rho(T)$ data set was smoothed with the adjacent-averaging methods with a 2 points of window before derivative was taken. (b) d$M$/d$H$ versus $H$ with the arrow denoting the metamagnetic-like feature seen in $M(H)$. (c) d$\rho$/d$H$ at 0.5 K as a function of magnetic field, with the criterion for $T_N$ and the minimum $\left(\dfrac{d\rho}{dH}\right)_{min}$. (d) $C_p/T$ as a function of temperature, arrows indicate a peak associated with the magnetic ordering and the maximum in the $C_p/T$ that emerges at $H\geq$47.5 kOe.}
\end{figure*}

\begin{figure}[tb]\centering
\includegraphics[width=1\linewidth]{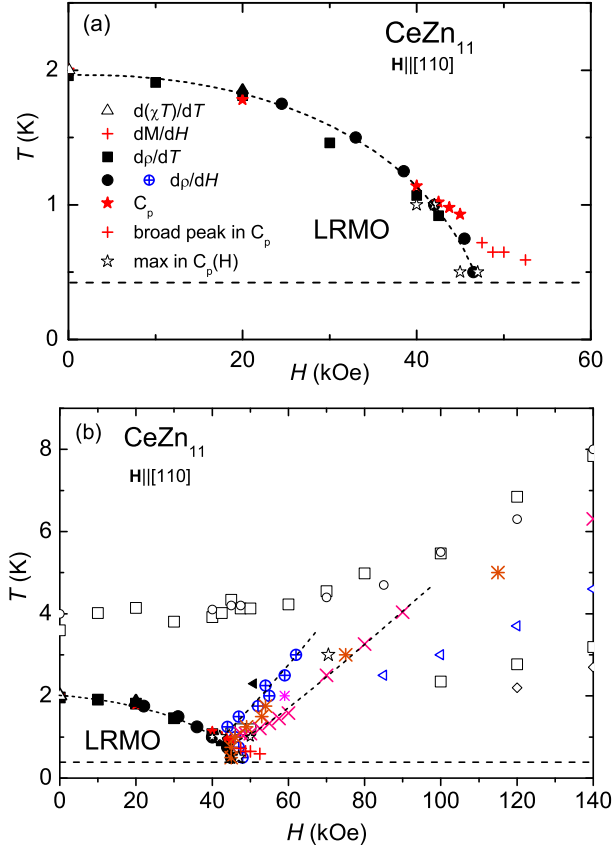}
\caption{\footnotesize (Color online) (a) and (b) Phase diagram, $\it H-T$, of CeZn$_{11}$ for $\bf{H}\|$[110]. Long-range magnetic order (LMRO) region is marked on the phase diagram. Legends: $\blacktriangle$ d$M$/d$H$, $\triangle$ d($\chi T$/d$T$), $\blacksquare$ and $\Box$ d$\rho$/d$T$ ( $T_N$ and broad maximum respectively), $\bullet$ and $\bigoplus$ d$\rho$/d$H$ ( $T_N$ and minimum respectively), $\star$ $C_p$, \textbf{+} broad peak in $C_p$, $\times$ max in $C_p$/$T$, open stars - features seen in $C_p(H)$, $\ast$ maximum in $\Delta\rho_H$, $\lhd$ $S_{min}$, $\circ$ $S_{\pm}$, $\Diamond$ $S_{\mp}$, and $\blacktriangleleft S(H)$. Dashed lines that run through the  data points are guides to the eye.  Dashed line at 0.4 K is the temperature limit of our measurements.}
\end{figure}

\begin{figure}[tb]
\centering
\includegraphics[width=1\linewidth]{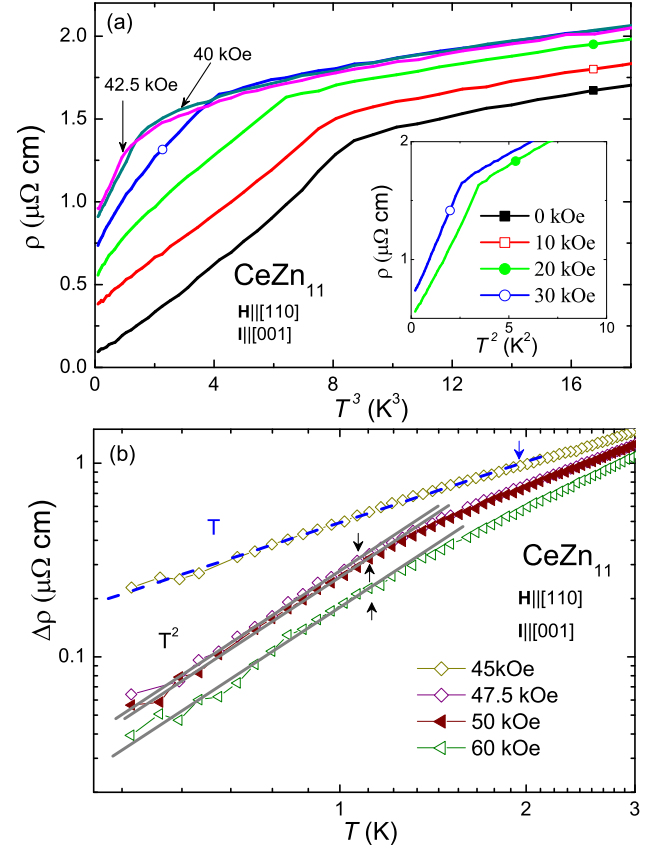}
\caption{\footnotesize (Color online) (a) $\rho(T)$ curves for CeZn$_{11}$ taken at different applied fields for $\bf{H}\|$[110] as a function of $T^3$ (main graph) and $T^2$ (inset). (b) Log-log plot of $\Delta\rho(T)$ for CeZn$_{11}$, 45 kOe$\leq H\leq$60 kOe. Dashed and solid lines represent $T$ and $T^2$ dependence respectively. Arrows denote the temperatures were the data deviate from linear and quadratic dependencies.} 
\end{figure}

\begin{figure}[tb]
\centering
\includegraphics[width=1\linewidth]{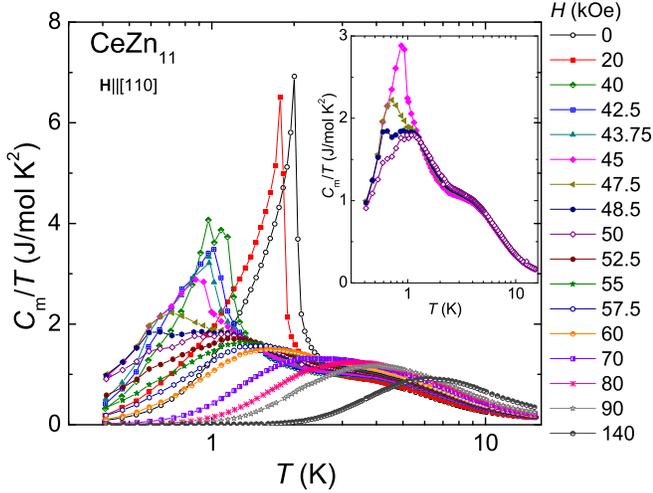}
\caption{\footnotesize (Color online) Semi-log plot of the magnetic part [$C_m$=$C_p$(CeZn$_{11}$)-$C_p$(LaZn$_{11}$)] of the heat capacity, $C_m/T$, as a function of $T$. Inset shows the data sets for 45 kOe$\leq H\leq$50 kOe.}
\end{figure}

\begin{figure}[tb]
\centering
\includegraphics[width=1\linewidth]{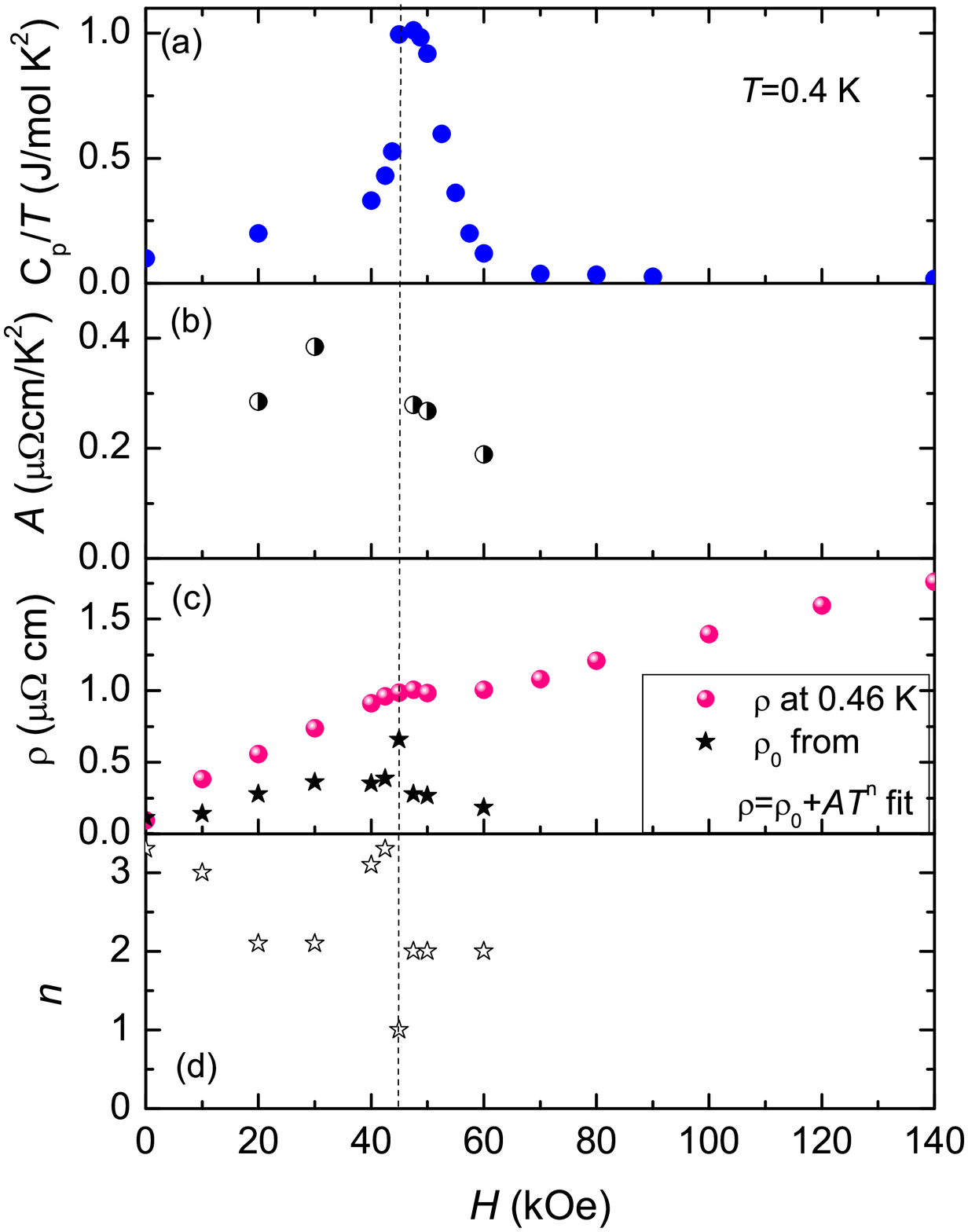}
\caption{\footnotesize (Color online) (a) $C/T$ at 0.4 K, (b) coefficient of the $T^2$ low fit of the resistivity, $A$, and (c) resistivity at 0.46 K and $\rho_0$, obtained from $\rho=\rho_0+AT^n$ fit, as a function of applied magnetic field of CeZn$_{11}$, (d) the exponent $n$ from $\rho=\rho_0+AT^n$ fit. }
\end{figure}

\section{Discussion}

CeZn$_{11}$ orders antiferromagnetically near 2 K. The zero-field temperature-dependent resistivity shows a broad shoulder, characteristic of that of Kondo compounds. The close proximity of the Schottky anomaly to the low AFM transition makes it hard to obtain or estimate the Sommerfeld coefficient $\gamma$ precisely and as a result to tell to what extend the density of states at the Fermi level is enhanced. According to Ref. \cite{Desgranges1982}, at the characteristic temperature $T_K$, the entropy reached by a Kondo system is 0.68$R$ ln2. For CeZn$_{11}$, 0.68$R$ ln2 of entropy is reached at $\sim$2 K meaning that $T_K<T_N$, possibly much less. The Kondo temperature can also be estimated from the paramagnetic Curie-Weiss temperature as $T_K=|\Theta_p|$/4.\cite{Gruner1974} From our Curie-Weiss law fit of the polycrystalline average (see Table 1) $\Theta_p$=-5.6$\pm$1.0 K which translates into $T_K$=1.4$\pm$1.0 K, which agrees with the $T_K$ obtained from the entropy. These estimates, taken together with our other data indicate that $T_N>T_K$ and possibly $T_N\gg T_K$.

Changing the applied magnetic field suppresses the AFM transition temperature of CeZn$_{11}$ and allows us to map the $T-H$ phase diagram and possibly find quantum critical point (QCP) effects. The criteria that were used to map the $T-H$ phase diagram are illustrated in Fig. 18. The criteria for inferring the AFM transition temperature, $T_N$, from the specific heat, $C_p$, and derivatives of resistivity, d$\rho$/d$T$ \cite{Fisher1968}, and magnetic susceptibility, d($\chi T$/d$T$) \cite{Fisher1962}, data are shown in Fig. 18(a). $T_N$ is denoted by an arrow. In d$\rho$/d$T$ data ($\rho(T)$ data were smoothed with the adjacent-averaging method with a 2 points of window before derivative was taken), in addition to the sharp peak at 1.93$\pm$0.07 K, that corresponds to $T_N$, another broad shoulder is evident at 3.6$\pm$0.5 K (denoted by an arrow in Fig. 18(a)). This broad feature moves to the higher temperature as the applied field is increased. As was aforementioned, the broad feature seen in the 1.85 K $M(H)$ data, centered at 18 kOe, which is associated with the suppression of the $T_N$ to below 1.85 K, can be clearly seen in d$M$/d$H$ plot in Fig. 18(b). Figure 18(c) illustrates features that were observed in d$\rho$/d$H$: the criterion for $T_N$ and a broadening minimum, the progression of which with increase of the temperature was marked with solid dots in Fig. 10.   

In the heat capacity measurements, as the applied magnetic field is increased, the sharp peak corresponding to the AFM transition moves to lower temperatures and broadens. For $H\geq$47.5 kOe, in addition to the broad maximum at low temperatures, another broad peak at slightly higher temperature develops and is more clearly seen in the $C_p/T$ versus $T$ plots, Fig. 18(d). This peak becomes broader and moves to the higher temperatures as the magnetic field is further increased. It has to be noted, that similar feature in the heat capacity was observed for YbNiSi$_3$ compound \cite{Budko2007} and was attributed to splitting by the applied field of the nearly degenerate crystal-electric field levels that form the zero-field-ground state. 

In Fig. 19(a), we present the low-temperature/low-field phase diagram for CeZn$_{11}$, $\bf{H}\|$[110], where, as the applied magnetic field is increased, the AFM order is suppressed below our base temperature of 0.4 K in the 45-47.5 kOe field range and a system is driven to a low-temperature state with no apparent long-range magnetic order. The broad shoulder seen in $C_p(T)$ above 45 kOe is shown as "+" and does seem to follow from the peak in $C_p$ associated with the $T_N$. Figure 19(a) looks promising in terms of bringing the AFM transition temperature to zero with the modest applied magnetic field. 

Figure 19(b) shows the $T-H$ phase diagram over a wider $T$ and $H$ range, where, in addition to the long-range magnetic order, we also added the rest of the features seen in the resistivity, heat capacity, Hall coefficient and TEP measurements. The broad Schottky-like features seen in $C_p/T$, the broad minimum observed in  d$\rho$/d$H$ and the broad maximum in $\rho_H$ appear to delineate a broad stripe that starts near the end of the AFM dome and around $H\sim$ 80 kOe starts to move to higher temperature with increasing field along with  $S_{min}$ and a feature seen in d$\rho$/d$T$. The origin of these features is probably due to the fact that different measurements see different degrees of scattering off of saturating moments and the CEF levels. The dashed lines that run through the data points obtained from d$\rho$/d$H$ and max in $C_p/T$ are the guides to the eye.

The evolution of the TEP sign change as the system is tuned by the applied magnetic field is also shown in Fig. 19(b) (open circles and diamonds). The TEP is negative in between open circles and open diamonds. We would like to point out, that for the TEP set-up used, 2.1 K was the lowest temperature reached.  Interestingly, the change of the sign of the TEP, $S_{\pm}$ (open circles), coincides with the feature seen in the d$\rho$/d$T_{max}$ (open squares).

In order to look for the Landau-Fermi-liquid or non-Landau-Fermi-liquid behavior as the AFM transition is suppressed, we further analysed the resistivity data by performing the fit of the low temperature part of the resistivity in the form $\rho=\rho_0+AT^n$, where $A$ is the coefficient and $n$ is the exponent. For the Landau-Fermi-liquid regime, the resistivity is governed by the electron-electron scattering, $A$ is the quasiparticle scattering amplitude and $n$=2. When the system is close to the AFM QCP, some theories predict $n\leq$5/3.\cite{Stewart2001} In the AFM state, the fit results in $n\approx $3 (Fig. 20(a)), except for the applied fields of 20 and 30 kOe for which $n=$2 (inset of Fig. 20(a)). Figure 20(b) shows log-log plot of $\Delta\rho=\rho-\rho_0=AT^n$ vs $T$ for 45 kOe$\leq H\leq$ 60 kOe, $\bf{H}\|$[110]. The low temperature $\rho(T)$ functional dependence for $H$=45 kOe is linear, $\rho(T)\sim T$, from base temperature of 0.46 K to 1.96 K. For the applied fields between 47.5 kOe and 60 kOe, the fit results in $n=$2 marking the region of the Landau-Fermi-liquid behavior. Given that the Landau-Fermi-liquid behavior holds only for a very small temperature range: from $\sim$1.12 K down to the base temperature of 0.46 K, further resistivity measurements for $T<$0.46 K will be required to have a larger span of the temperatures for which the Landau-Fermi-liquid regime holds. We would like to point out that, given the very large RRR value of CeZn$_{11}$, normal metal contribution to magneto-resistance may obscure the functional dependence of resistivity associated with strong electronic correlations (see inset to Fig. 9). 

To look for non-Landau-Fermi-liquid behavior in the heat capacity data of CeZn$_{11}$, the magnetic part of the heat capacity, $C_m/T$, as a function of temperature  is plotted on a semi-log plot in Fig. 21. $C_m$ was obtained by subtracting $C_p$(LaZn$_{11}$) ($H$=0) using the assumption that for LaZn$_{11}$ $C_p(T)$ is essentially field independent. For a classical field induced quantum critical system, for $H=H_c$ there should be a region of the logarithmic divergence, $C_m/T\propto$-ln $\it T$, indicating a non-Landau-Fermi-liquid regime. The inset to Fig. 20 shows data for 45 kOe$\leq H\leq$ 50 kOe and no such non-Landau-Fermi-liquid like signature is found.

Figure 21 also clearly shows that for $H\leq$60 kOe the electronic specific heat coefficient $\gamma$, that reflects the effective mass of the 4$f$ electrons, clearly becomes very small. This is in contrast to such QCP systems such as YbAgGe \cite{Budko2004} or YbPtBi\cite{Mun2013}, where $\gamma$ stays significantly enhanced for wide field ranges above $H_c$. Given our 0.4 K base temperature, we can plot $C_p/T$ at 0.4 K as a function of applied field as shown in Fig. 22(a). As the magnetic field is increased, $C_p/T$ increases, reaches the highest value of 1 J/(mol K) at 47.5 kOe and then steadily and rapidly decreases to $\sim$18 mJ/(mol K) at $H$=140 kOe. The coefficient of the $T^2$ low temperature fit of the resistivity, $A$, as a function of the applied field is shown in Fig. 22(b) and seems to have a weak maximum at the same field where $C_p/T|_{0.4 K}$ has a maximum. Figure 22(c) shows the evolution with field of the resistivity, $\rho$, at 0.46 K and $\rho_0$, obtained from $\rho=\rho_0+AT^n$ fit. As the applied field is increased, both $\rho$ and $\rho_0$ show features at about $H$=45 kOe. $\rho$ at 0.46 K increases, reaches a small maximum at 47.5 kOe then goes through a shallow minimum and increases again. Much clearer feature is seen in $\rho_0$, where, as the magnetic field is increased, $\rho_0$ increases until at $H$=45 kOe where there is a sudden increase followed at higher fields by a sharp drop and then steady decreases as the magnetic field is being further increased. Figure 22(d) shows the exponent $n$ from $\rho=\rho_0+AT^n$ fit. Taken as a whole, Fig. 22 is more consistent with a quantum phase transition similar to Ref.\cite{Budko2007,Wu} that does not manifest any critical behavior. It becomes likely that CeZn$_{11}$ manifest a simple, local moment-like, metamagnetic phase transition as a function of field.

\section{Conclusion}
We have studied the electrical, magnetic, and thermal properties of single crystals of CeZn$_{11}$ by the means of magnetization, resistivity, specific heat, Hall coefficient and thermoelectric power. Based on the analysis of our measurements, CeZn$_{11}$ may be classified as a mildly correlated, local moment system with $T_K<T_N\sim$2 K. Rather low CEF energy level splitting is influencing the transport and thermodynamic properties of the compound. CeZn$_{11}$ manifests a strong anisotropy with the [110] direction being the easy axis. The Hall coefficient is constant at high temperatures followed by a sign reversal at low temperatures. Thermoelectric power shows an almost linear temperature dependence at high temperatures and reverses a sign below 4 K at zero applied magnetic field. Both Hall resistivity and thermoelectric power are positive at high temperatures indicating hole-type carriers dominating the transport properties of CeZn$_{11}$. The constructed $T-H$ phase diagram indicates that the applied magnetic field drives the AFM order temperature below 0.4 K, our lowest temperature measured, by $H\sim$45-47.5 kOe for $\bf{H}\|$[110]. For the easy axis, $\bf{H}\|$[110], the linear behavior in the $\rho(T)$ was observed only for $H$=45 kOe at 0.46 K$\leq T\leq$1.96 K followed by the Landau-Fermi-liquid regime for the limited range of applied magnetic field, 47.5 kOe$\leq H\leq$60 kOe. No non-Landau-Fermi liquid behavior was observed in the heat capacity data.  Most likely, as the AFM transition is suppressed, for the magnetic field applied along the easy [110] axis, we observe a quantum phase transition, maybe even of the first order, as is the case for local moment rare earth meta-magnetism, rather than a field-induced quantum critical point.

\section{acknowledgments}
The authors would like to thank A. I. Goldman, A. Jesche, H. Kim and T. Kong for insightful discussions. This work was done at Ames Laboratory, US DOE, under contract $\#$DE-AC02-07CH111358. This work was supported by the US Department of Energy, Office of Basic Energy Science, Division of Materials Sciences and Engineering. X. Lin and V. Taufour are supported by AFOSR-MURI grant No. FA9550-09-1-0603.

\section*{Appendix  }
\section*{Analysis of the dHvA oscillation of LaZn$_{11}$ }

To measure the dHvA oscillations in LaZn$_{11}$ the applied field was varied in the constant intervals of 1/$H$ in 37-70 kOe field range for $\bf{H}\|$[110] and $\bf{H}\|$[001]. To separate quantum oscillations from observed data,  the linear background magnetization was subtracted. The resultant magnetization as a function of inverse field is plotted in Fig. 23. The fast Fourier transforms (FFTs) of these data are shown in Fig. 24. The frequencies of the dHvA oscillations for $\bf{H}\|$[110] are smaller than the ones observed for $\bf{H}\|$[001]. For, $\bf{H}\|$[110] three large amplitude peaks are present at 0.23 MG ($\alpha$), 2.51 MG ($\gamma)$, and 3.21 MG ($\delta)$ and a smaller amplitude peak at 0.47 MG ($\beta$). For, $\bf{H}\|$[001] a large amplitude peak is present at 6.32 MG ($\delta_1$) and smaller amplitude peaks at 0.32 MG ($\alpha_1$), 0.55 MG ($\beta_1)$, 4.34 MG ($\gamma_1)$, and 12.63 MG ($\varepsilon_1$). Frequencies of $\beta$ and $\beta_1$ are almost the same. The frequencies for LaZn$_{11}$ obtained from the FFT are much smaller then the ones obtained for CeZn$_{11}$ (see inset in Fig. 17).

\begin{figure}[tbh]
\centering
\includegraphics[width=1\linewidth]{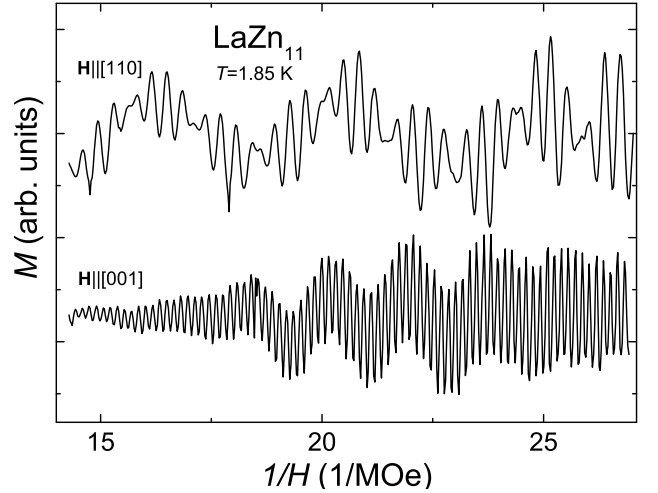}
\caption{\footnotesize Magnetization isotherms of LaZn$_{11}$ for $\bf{H}\|$[110] and $\bf{H}\|$[001] at $T$=1.85 K plotted versus 1/$H$. }
\end{figure}

The effective masses calculated from the temperature-dependence of the FFT amplitudes, $A$, of the oscillations can be used to determine the effective masses of the orbits with the help of the Lifshitz-Kosevitch equation \cite{Shoenberg1984}:
\begin{multline*}
M=-2.602\times 10^{-6}\left(\frac{2\pi}{HA^{''}}\right)^{1/2}\times 
\\
\frac{GFT\exp(-\alpha px/H)}{p^{3/2}\sinh(-\alpha pT/H)}\sin\left[\left(\frac{2\pi pF}{H}\right)-\frac{1}{2}\pm \frac{\pi}{4}\right],
\end{multline*}
where $\alpha$=1.47(m/m$_0$)$\times 10^5$ G/K, $A^{''}$ is the second derivative of the cross sectional area of the Fermi surface with respect to wave vector along the direction of the applied field, $G$ is the reduction factor arising from electron spin, $F$ is the frequency of an orbit, $p$ is the number of the harmonic of the oscillation, and $x$ is the Dingle temperature.

From the slope of ln($A/T$) plotted as a function of temperature, inset of Fig. 24, the effective masses for LaZn$_{11}$ for $\bf{H}\|$[110] were found to be $m_{\alpha}=0.11(1)m_0$, $m_{\beta}=0.18(1)m_0$, $m_{\gamma}=0.20(1)m_0$, and $m_{\delta}=0.11(1)m_0$, where $m_0$ is the bare electron mass.

\begin{figure}[tbh]
\centering
\includegraphics[width=1\linewidth]{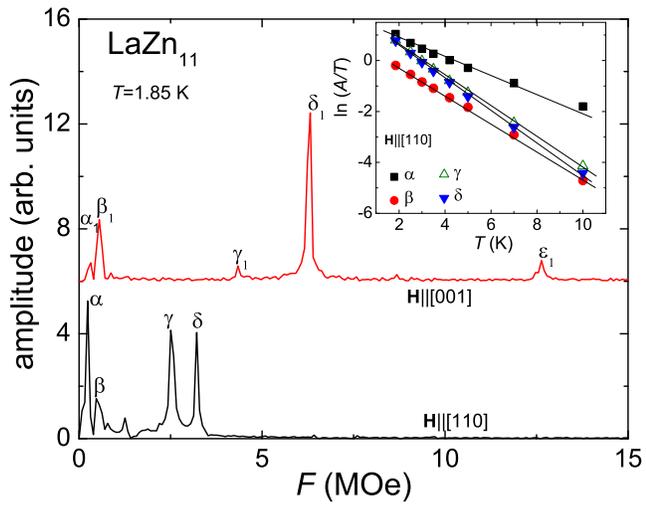}
\caption{\footnotesize (Color online) FFT spectra of the dHvA data of LaZn$_{11}$ for $\bf{H}\|$[110] and $\bf{H}\|$[001]. Inset: temperature dependence of the FFT amplitudes, $A$, of the observed oscillations for $\bf{H}\|$[110]. }
\end{figure}

\end{document}